\documentclass[10pt]{article}
\usepackage{multicol}
\usepackage{graphicx}
\usepackage{amsmath}

\begin{document}

\begin{center}
{\Large\bf Event patterns extracted from top quark-related spectra
in proton-proton collisions at 8 TeV}

\vskip0.75cm

Ya-Hui Chen$^1$, Fu-Hu Liu$^{1,}${\footnote{E-mail:
fuhuliu@163.com; fuhuliu@sxu.edu.cn}}, Roy A. Lacey$^2$

{\small\it $^1$Institute of Theoretical Physics \& State Key
Laboratory of Quantum Optics and Quantum Optics Devices,

Shanxi University, Taiyuan, Shanxi 030006, China

$^2$Departments of Chemistry \& Physics, Stony Brook University,
Stony Brook, NY 11794, USA}
\end{center}

\vskip0.5cm

{\bf Abstract:} We analyze the transverse momentum ($p_T$) and
rapidity ($y$) spectra of top quark pairs, hadronic top quarks,
and top quarks produced in proton-proton ($pp$) collisions at
center-of-mass energy $\sqrt{s}=8$ TeV. For $p_T$ spectra, we use
the superposition of the inverse power-law suggested by the QCD
(quantum chromodynamics) calculus and the Erlang distribution
resulting from a multisource thermal model. For $y$ spectra, we use
the two-component Gaussian function resulting from the revised
Landau hydrodynamic model. The modelling results are in agreement
with the experimental data measured at the detector level, in the
fiducial phase-space, and in the full phase-space by the ATLAS
Collaboration at the Large Hadron Collider (LHC). Based on the
parameter values extracted from $p_T$ and $y$ spectra, the event
patterns in three-dimensional velocity
($\beta_x$-$\beta_y$-$\beta_z$), momentum ($p_x$-$p_y$-$p_z$), and
rapidity ($y_1$-$y_2$-$y$) spaces are obtained, and the
probability distributions of these components are also obtained.
\\

{\bf Keywords:} top quark-related spectra, event pattern,
three-dimensional space
\\

{\bf PACS:} 14.65.Ha, 25.75.Ag, 25.75.Dw, 24.10.Pa

\vskip1.0cm

\begin{multicols}{2}

{\section{Introduction}}

The top quark is the heaviest particle in the standard model, and is
very different from the other quarks. It is expected that the top
quark may have some special characteristics and be related to
new physics beyond the standard model. Therefore, it is very
important to study its characteristics more. The top quark was
first found by the Tevatron at the Fermi National Accelerator
Laboratory [1--4], and the successful operation of the CERN Large
Hadron Collider (LHC) has brought the study of the top quark into a
more precise measurement period. High energy collisions are the
only way to study the top quark in experiments. Generally, due to
the complexity of the process in an extremely short time,
theoretical physicists need to use models to analyze the
properties of observables, instead of studying the interacting
systems directly.

Among the kinematic observables, the transverse momentum ($p_T$)
and rapidity ($y$) are always hot topics for theoretical
physicists. Some phenomenological models and formulas have been
proposed to fit the $p_T$ and $y$ spectra. For $p_T$ spectra, many
formulas can be used, such as the standard (Fermi-Dirac,
Bose-Einstein, or Boltzmann) distribution [5--8], Tsallis
statistics [8--14], the inverse power-law [15--17], the Erlang
distribution [18], the Schwinger mechanism [19--22], and
combinations of these, while $y$ spectra can be described by the
one-, two-, or three-component Gaussian function. Except for these
analytic expressions, many models based on the Monte Carlo method
have been used to find arithmetic solutions of the spectra of
$p_T$ and $y$, and other interesting results which contain, but
are not limited to, chemical and kinetic freeze-out temperatures,
chemical potential, transverse flow velocity, particle ratio, and
so forth.

High energy collisions are complex processes in which the
production mechanisms of different types of particles are
different. In particular, at the same stage of a collision
process, different types of particles may undergo different ways
of propagation. Although some spectra of different types of
particles can be described or fitted by particular theoretical
models or functions, we are interested in the differences in their
production. For example, before leaving the interacting region,
most light flavor particles undergo the stage of kinetic
freeze-out and local thermal equilibrium. Heavy flavor particles
do not undergo this stage. It is hard to learn more information
from the limited spectra available in experiments. We hope to use
a simple method to extract some intuitive pictures from the
limited spectra, so that some differences in the production of
different types of particles can be observed. These differences
are useful in understanding the production mechanisms of different
types of particles.

In order to understand the sophisticated collision process and
mechanism intuitively, we can use the method of event pattern
(particle scatter plot) at the last stage of particle production
to obtain information about the interacting system. Using this
method, we have analyzed the scatter plots of net-baryons produced
in central gold-gold (Au-Au) collisions at BNL Relativistic Heavy
Ion Collider (RHIC) energies in three-dimensional momentum
($p_x$-$p_y$-$p_z$) space, three-dimensional momentum-rapidity
($p_x$-$p_y$-$y$) space, and three-dimensional velocity
($\beta_x$-$\beta_y$-$\beta_z$) space [23]; charged particles
produced in proton-proton ($pp$) and lead-lead (Pb-Pb) collisions
at 2.76 TeV (one of the LHC energies) in three-dimensional
$p_x$-$p_y$-$p_z$ space and $\beta_x$-$\beta_y$-$\beta_z$ space
[24]; as well as $Z$ bosons and quarkonium states produced in $pp$
and Pb-Pb collisions at LHC energies in two-dimensional transverse
momentum-rapidity ($p_T$-$y$) space and three-dimensional
$\beta_x$-$\beta_y$-$\beta_z$ space [25]. Due to the variety
of produced particles, further analyses of the event patterns of
other particles such as the top quarks are needed.

In this paper, we mainly use the superposition of the inverse
power-law suggested by the QCD (quantum chromodynamics) calculus
[15--17] and the Erlang distribution resulting from a multisource
thermal model [18], and the two-component Gaussian function
resulting from the revised Landau hydrodynamic model [26--29] to
fit the $p_T$ and $y$ spectra of the top quark-related products
produced in $pp$ collisions at the center-of-mass energy
$\sqrt{s}=8$ TeV measured at the detector level, in the fiducial
phase-space, and in the full phase-space by the ATLAS
Collaboration at the LHC [30]. The related parameters can be
extracted from the fitting. Based on the parameters and using the
Monte Carlo method, we can obtain the event patterns in
three-dimensional $\beta_x$-$\beta_y$-$\beta_z$,
$p_x$-$p_y$-$p_z$, and rapidity ($y_1$-$y_2$-$y$) spaces. The
probability distributions of these components can also be obtained.

The remainder of this paper is structured as follows. A brief
description of the model and method is given in Section 2. Then,
the results and discussion are presented in Section 3. Finally, we
summarize our main observations and conclusions in Section 4.
\\

{\section{Model and method}}

As the heaviest particle in the standard model, the formation of
the top quark is expected to be through the hard scattering process
among partons (quarks and gluons) with high energy. The
top quark-related $p_T$ spectra have a very wide range of
distributions. This means that in some cases the spectra can in
fact be divided into two parts. One part is in the relatively high $p_T$
region and mainly contributed by the real hard (the harder)
scattering process, and the other part is in the relatively low $p_T$
process and mainly contributed by the not too hard (the hard)
scattering process.

For the harder and hard processes, we have to choose a
superposition distribution which has two components to describe
the $p_T$ spectra. Of the two components, one is for the harder
scattering process and the other for the hard scattering
process. The relative contributions of the harder scattering
process are expected to be different for different products such
as top quark pairs ($t\bar t$ systems), hadronic top
(hadronic $t$) quarks (hadronically decaying top quarks), and
 top ($t$) quarks (semileptonically and hadronically
decaying top quarks).

For the harder scattering process, we can use the inverse
power-law which results from the QCD calculus [15--17] in high
energy collisions to describe the $p_T$ spectra. For the hard
scattering process, we can use the Erlang distribution which
results from a multisource thermal model [18]. Although the Erlang
distribution is not sure to be the best choice for the hard
scattering process, it is a good one to fit many data. In fact,
the Erlang distribution is also used for the soft excitation
process. The inverse power-law plays a significant role in the
region of relatively high $p_T$, and the Erlang distribution
contributes mainly in the region of relatively low $p_T$.

According to the QCD calculus [15--17], we have the inverse
power-law in the form
\begin{equation}
f_1(p_T)=Ap_T \bigg(1+\frac{p_T}{p_0} \bigg)^{-n},
\end{equation}
where $A$ denotes the normalization constant which makes
$\int_0^{\infty} f_1(p_T) dp_T=1$, and $p_0$ and $n$ are free
parameters and influence the value of $A$.

According to the multisource thermal model [18], the spectra of
$p_T$ for a given set of data selected in a special condition can
be described by the Erlang distribution
\begin{equation}
f_2(p_{T})=\frac{p_T^{m-1}}{(m-1)!\langle p_{Ti} \rangle^m} \exp
\bigg(- \frac{p_{T}}{\langle p_{Ti} \rangle} \bigg),
\end{equation}
where $\langle p_{Ti} \rangle$ and $m$ are free parameters. In
particular, $\langle p_{Ti} \rangle$ denotes the average value of
$p_{Ti}$, where $i=1$ to $m$, and $m$ denotes the number of
contribution sources, which are in fact the participant partons,
which contribute the same exponential function to $p_T$.
Generally, $m=2$ or 3 due to only two or three partons taking part
in the formation of each particle.

The top quark-related $p_T$ spectrum is a superposition of the
inverse power-law and the Erlang distribution. Let $k$ denote the
relative contribution of the inverse power-law. Then, the relative
contribution of the Erlang distribution is $1-k$. We have the
normalized distribution
\begin{equation}
f_0(p_{T})=kf_{1}(p_{T})+(1-k)f_{2}(p_{T}).
\end{equation}
In many cases, the spectra of $p_T$ in experiments are presented
in terms of non-normalized distributions. To give a comparison with
the experimental data, the normalized constant ($N_{p_T}$) is
needed.

According to the Landau hydrodynamic model and its revisions
[26--29], the $y$ spectrum is a Gaussian function [28,
29]
\begin{equation}
f_y(y)=\frac{1}{\sqrt{2\pi} \sigma_y} \exp \bigg[-
\frac{(y-y_C)^2}{2\sigma_y^2} \bigg],
\end{equation}
where $y_C$ denotes the mid-rapidity (peak position) and
$\sigma_y$ denotes the distribution width. In the center-of-mass
reference frame, $y_C=0$ corresponds to symmetric collisions
such as the $pp$ collisions considered in the present work.

In many cases, the Gaussian function cannot describe the $y$
spectra very well, and we need at least two Gaussian functions for the
$y$ spectrum. That is,
\begin{align}
f_y(y) &= \frac{k_B}{\sqrt{2\pi} \sigma_{y_B}} \exp \bigg[-
\frac{(y-y_B)^2}{2\sigma_{y_B}^2} \bigg] \nonumber\\
&+ \frac{1-k_B}{\sqrt{2\pi} \sigma_{y_F}} \exp \bigg[-
\frac{(y-y_F)^2}{2\sigma_{y_F}^2} \bigg],
\end{align}
where $k_B$ ($1-k_B$), $y_B$ ($y_F$), and $\sigma_{y_B}$
($\sigma_{y_F}$) denote respectively the relative contribution
ratio, peak position, and distribution width of the first (second)
component, distributed in the backward (forward) rapidity
region. Due to the symmetry of $pp$ collisions, we have
$k_B=1-k_B=0.5$, $y_B=-y_F$, and $\sigma_{y_B}=\sigma_{y_F}$. When
comparing with experimental data, the normalization constant
($N_y$) is needed.

In the present work, we use the Monte Carlo method to get
discrete values which are used in the event patterns in
three-dimensional $\beta_x$-$\beta_y$-$\beta_z$,
$p_x$-$p_y$-$p_z$, and $y_1$-$y_2$-$y$ spaces. Let $R$, $r_i$, and
$R_{1-5}$ denote random numbers distributed evenly in $[0,1]$. To
get the discrete values, the variable $p_T$ in Eq. (1) or in the
first component in Eq. (3) obeys the following formula
\begin{equation}
\int_0^{p_{T}}f_{1}(p_{T})dp_{T} <R
<\int_0^{p_{T}+dp_{T}}f_{1}(p_{T}) dp_{T}.
\end{equation}
The variable $p_T$ in Eq. (2) or in the second component in Eq.
(3) is obtained by
\begin{equation}
p_T=-\langle p_{Ti} \rangle \sum_{i=1}^{m} \ln r_i = -\langle
p_{Ti} \rangle \ln \prod_{i=1}^m r_i.
\end{equation}
As for the variable $y$ in the first and second components in Eq.
5, we have
\begin{equation}
y=\sigma_y \sqrt{-2\ln R_1} \cos(2\pi R_2) +y_B
\end{equation}
and
\begin{equation}
y=\sigma_y \sqrt{-2\ln R_3} \cos(2\pi R_4) +y_F
\end{equation}
respectively.

An isotropic emission in the transverse plane results in the
azimuthal angle $\varphi$ being
\begin{equation}
\varphi=2\pi R_5
\end{equation}
which is distributed evenly in $[0,2\pi]$. The momentum components
are
\begin{equation}
p_{x}=p_{T}\cos \varphi,
\end{equation}
\begin{equation}
p_{y}=p_{T}\sin \varphi,
\end{equation}
and
\begin{equation}
p_{z}=\sqrt{p_T^2+m_0^2}\sinh y,
\end{equation}
where $m_0$ is the peak mass in the invariant mass spectrum of the
$t\bar t$ systems [30], or the rest mass of the top quark in the case
of hadronic top quarks or top quarks.

The energy $E$ is
\begin{equation}
E=\sqrt{p_T^2+m_0^2}\cosh y.
\end{equation}
The velocity components are
\begin{equation}
\beta_{x}=\frac{p_{x}}{E},
\end{equation}
\begin{equation}
\beta_{y}=\frac{p_{y}}{E},
\end{equation}
and
\begin{equation}
\beta_{z}=\frac{p_{z}}{E}.
\end{equation}

For $y_1$ and $y_2$, we use the definitions
\begin{equation}
y_1=\frac{1}{2}\ln \bigg( \frac{E+p_{x}}{E-p_{x}} \bigg)
\end{equation}
and
\begin{equation}
y_2=\frac{1}{2}\ln \bigg( \frac{E+p_{y}}{E-p_{y}} \bigg)
\end{equation}
for the rapidities in the directions of the $ox$ and $oy$ axes
respectively. We get $y$ using  Eq. (8) or (9) directly.

In the concrete calculation, we need a few steps to generate the
event patterns, i.e. the three-dimensional distributions of
particle scatters.

i) We use Eq. (3) to fit the $p_T$ spectra and Eq. (5) to fit
the $y$ spectra so that the values of free parameters and
normalization constants can be obtained;

ii) According to the values of the free parameters obtained by the
fitting in the first step, we may use Eq. (6) or (7) on the basis
of $k$ to get a discrete value of $p_T$, and Eq. (8) or (9) with
an equal probability to get a discrete value of $y$;

iii) We use Eq. (10) to get a discrete value of $\varphi$, and
Eqs. (11)--(13) to get a set of discrete values of
($p_x,p_y,p_z$);

iv) We use Eq. (14) to get a discrete value of $E$, and Eqs.
(15)--(17) to get a set of discrete values of
($\beta_x,\beta_y,\beta_z$);

v) We use Eq. (18) to get a discrete value of $y_1$, Eq. (19)
to get a discrete value of $y_2$, and the discrete value of $y$ is
obtained directly from Eq. (8) or (9). Then, a set of discrete
values of ($y_1,y_2,y$) is obtained;

vi) Repeating steps ii) to v) 1000 times, we can obtain 1000
sets of ($p_x,p_y,p_z$), 1000 sets of ($\beta_x,\beta_y,\beta_z$),
and 1000 sets of ($y_1,y_2,y$);

vii) Finally, the three-dimensional distributions of particle
scatters are plotted in three three-dimensional spaces.

The probability distributions of the
various components are obtained by statistics.
\\

{\section{Results and discussion}}

Before describing the comparisons with experimental data, we
first introduce  the meanings of ``the detector level", ``the
fiducial phase-space'', and ``the full phase-space". According to
Ref. [30], ``the event
selection consists of a set of requirements based on the general
event quality and on the reconstructed objects", defined by
definite conditions, ``that characterize the final-state event
topology". Each requirement or condition for quantities of the
considered event has to be detected, identified, and selected by
various types of detectors, which is referred to as the detector level.
The fiducial phase space for the measurements presented in Ref.
[30] is defined ``using a series of requirements applied to
particle-level objects close to those used in the selection of the
detector-level objects". The full phase space for the measurements
presented in Ref. [30] is defined ``by the set of $t\bar t$ pairs
in which one top quark decays semileptonically (including $\tau$
leptons) and the other decays hadronically".

Although we intend to ``describe" separately the spectra of top
quark pairs, hadronic top quarks, and top quarks at the ``detector
level", in the ``fiducial" phase-space, and in the ``full"
phase-space, these concepts are not independent. In fact, the
differences between them arise due to the decay of the top quark
and also due to the finite coverage of the detector. Predicting
the relationship between them is one of the great successes of
perturbative QCD: the experiments only measure particle data at
the detector-level, which is then extrapolated with theoretical
tools to the level of top quarks and, if desired, can also be
corrected to the fiducial or full phase-space.

On the short-cut process of the rest mass for the $t\bar t$
systems in the cases of measuring the invariant mass spectra by
the three requirements which are (i) at the detector level, (ii)
in the fiducial phase-space, and (iii) in the full phase-space, we
take $m_0=463.2$, 382.5, and 372.5 GeV, respectively, as in Ref.
[30]. For the rest mass of (hadronic) top quark, we take
$m_0=172.5$ GeV, also from Ref.~[30]. It is very important to use
the correct rest mass in the extraction of event patterns. In the
case of giving the spectra of $p_T$ and $y$, the most important
issue is the rest mass.

Figure 1 shows the event yields for (a)(c) $p_T$ and (b)(d) $y$
of (a)(b) the $t\bar t$ systems and (c)(d) the hadronic top quarks
produced in $pp$ collisions at $\sqrt{s}=8$ TeV, where $y$ spectra
are presented in terms of absolute values. The symbols represent
the experimental data of the ATLAS Collaboration [30] measured in
the combined electron and muon selections at the detector level,
and the error bars are the combined statistical and systematic
uncertainties, where the integral luminosity corresponds to $20.3$
fb$^{-1}$. The solid curves are our results calculated by using
(a) the inverse power-law, (c) the superposition of the inverse
power-law and the Erlang distribution, and (b)(d) the
two-component Gaussian distribution, respectively. The dashed
curve in (a) is the result calculated by using the Erlang
distribution with $m=1$, which is in fact the exponential
distribution for the purpose of comparison. In the calculation, we
use the method of least squares to determine the values of
parameters. The values of free parameters [$p_0$, $n$, $k$,
$\langle p_{Ti} \rangle$, $y_F$ ($=-y_B$), and $\sigma_{y_F}$
($=\sigma_{y_B}$)], normalization constants ($N_{p_T}$ and
$N_{y}$), and $\chi^2$ per degree of freedom ($\chi^2$/dof) are
listed in Tables 1--3 for different sets of products and
parameters, where $k=1$ for the $t\bar t$ systems and $m=3$ for
the hadronic top quarks.

One can see from Fig. 1 and Tables 1--3 that the results
calculated by the hybrid model are in agreement with the
experimental $p_T$ and $y$ spectra of both the $t\bar t$ systems
and hadronic top quarks produced in $pp$ collisions at
$\sqrt{s}=8$ TeV measured at the detector level by the ATLAS
Collaboration at the LHC. In particular, the $t\bar t$ systems
show only the contribution of the harder scattering process (with
$k=1$), which means that $\langle p_{Ti} \rangle$ and $m$ for the
$t\bar t$ systems are not available. The hadronic top quarks show
mainly the contribution of the hard scattering process (with a small
$k$). This implies that more collision energy is needed to
create the $t\bar t$ system. As part of the $t\bar t$ system,
the hadronic top quark takes up part of the energy of the $t\bar t$
system, which results in a not too hard scattering process.

Figure 2 shows the fiducial phase-space normalized differential
cross-sections for (a)(b) the $t\bar t$ systems and (c)(d) the
hadronic top quarks, where $\sigma$ denotes the cross-section.
Figure 3 shows the full phase-space normalized differential
cross-sections for (a)(b) the $t\bar t$ systems and (c)(d) the top
quarks. The corresponding parameters are also listed in Tables
1--3 for different sets of products and parameters, where $k=1$
for the $t\bar t$ systems and $m=3$ for the hadronic top quarks
and top quarks, which are listed only in the captions of Tables 1
and 2, but not as a separate column in the tables.

One can also see from Figs. 2 and 3 and Tables 1--3 that the
modelling results are in agrement with the experimental data of
the fiducial phase-space normalized differential cross-sections
for the $t\bar t$ systems and the hadronic top quarks, and the
full phase-space normalized differential cross-sections for the
$t\bar t$ systems and the top quarks, produced in $pp$ collisions
at 8 TeV measured by the ATLAS Collaboration. In particular, for
the $t\bar t$ systems in both the fiducial and full phase-spaces,
we also have $k=1$. For both the hadronic top quarks in the
fiducial phase-space and the top quarks in the full phase-space,
we also have a small $k$.

\begin{figure*}
\hskip-1.0cm \begin{center}
\includegraphics[width=15.0cm]{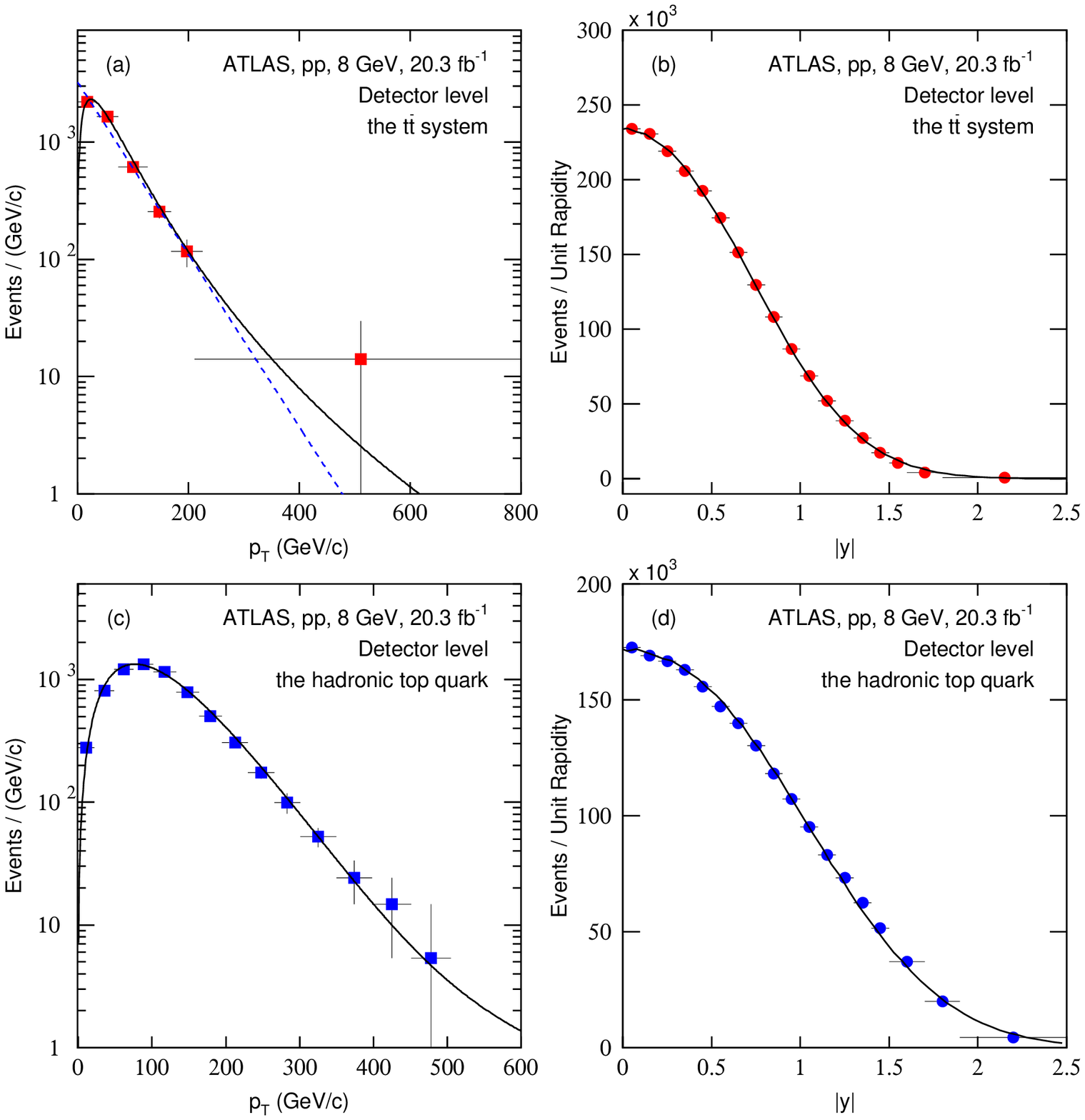}
\end{center}
\vskip0.5cm {\small Fig. 1. (a)(c) Transverse momentum and (b)(d)
rapidity spectra of (a)(b) the $t\bar t$ systems and (c)(d) the
hadronic top quarks produced in $pp$ collisions at $\sqrt{s}=8$
TeV, where the rapidity spectra are presented in terms of absolute
values. The symbols represent the experimental data of the ATLAS
Collaboration [30] measured in the combined electron and muon
selections at the detector level. The solid curves are our results
calculated by using the (a) inverse power-law, (c) superposition
of inverse power-law and Erlang distribution, and (b)(d)
two-component Gaussian distribution, respectively. The dashed
curve in (a) is the result calculated by using the Erlang
distribution with $m=1$, which is in fact the exponential
distribution.}
\end{figure*}

\begin{table*}
{\scriptsize Table 1. Values of free parameters ($p_{0}$ and $n$),
normalization constant ($N_{p_T}$), and $\chi^2$/dof corresponding
to the curves in Figures 1(a), 2(a), and 3(a), where $k=1$, which
means that $\langle p_{Ti} \rangle$ and $m$ are not available.
\begin{center}
\begin{tabular}{cccccc}
\hline\hline Figure & Type & $p_{0}$ (GeV/$c$) & $n$  & $N_{p_T}$ & $\chi^2$/dof \\
\hline
Figure 1(a) & $t\bar t$ & $162.86\pm4.20$ & $7.70\pm0.40$ & $(1.93\pm0.05)\times10^{5}$ & 2.88 \\
Figure 2(a) & $t\bar t$ & $52.73\pm2.60$ & $4.34\pm0.20$ & $1.00$ & 4.26 \\
Figure 3(a) & $t\bar t$ & $45.83\pm2.50$ & $4.30\pm0.20$ & $1.00$ & 1.43 \\
\hline
\end{tabular}%
\end{center}}
\end{table*}

\begin{figure*}
\hskip-1.0cm \begin{center}
\includegraphics[width=15.0cm]{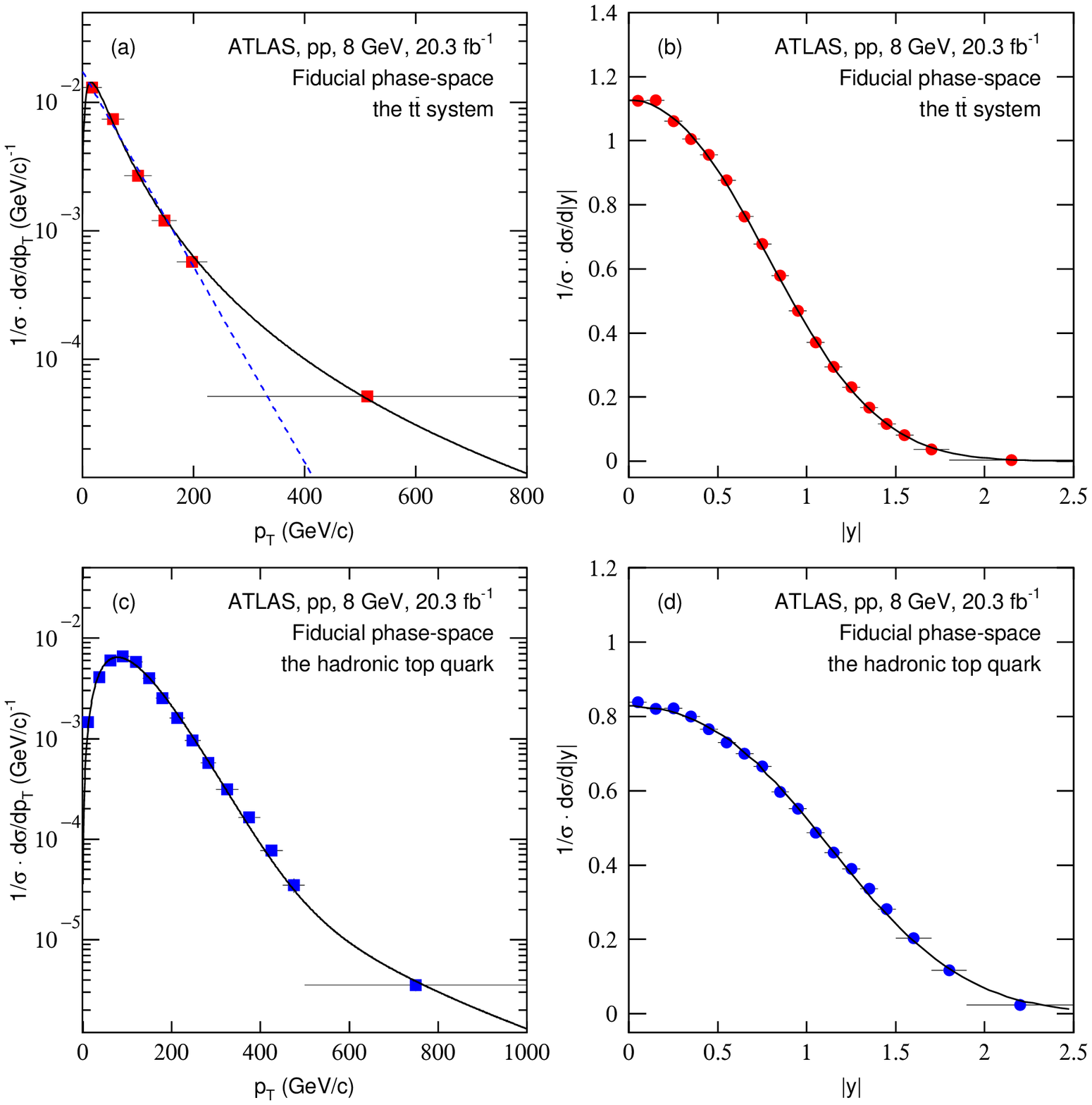}
\end{center}
\vskip0.5cm {\small Fig. 2. (a)(c) Transverse momentum and (b)(d)
rapidity normalized differential cross-sections of (a)(b) the
$t\bar t$ systems and (c)(d) the hadronic top quarks produced in
$pp$ collisions at $\sqrt{s}=8$ TeV. The symbols represent the
experimental data of the ATLAS Collaboration [30] measured in the
combined electron and muon selections at the fiducial phase-space
level. The solid curves are our results calculated by using the
(a) inverse power-law, (c) superposition of inverse power-law and
Erlang distribution, and (b)(d) two-component Gaussian
distribution, respectively. The dashed curve in (a) is the result
calculated by using the Erlang distribution with $m=1$, which is
in fact the exponential distribution. }
\\
\end{figure*}

\begin{table*}
{\scriptsize Table 2. Values of free parameters ($p_{0}$, $n$,
$k$, $\langle p_{Ti} \rangle$, and $m$), normalization constant
($N_{p_T}$), and $\chi^2$/dof corresponding to the curves in
Figures 1(c), 2(c), and 3(c), where the values of $m$ in the
Erlang distribution are invariably taken to be 3 and are not
listed in a separate column.
\begin{center}
\begin{tabular}{cccccccc}
\hline\hline Figure & Type & $p_{0}$ (GeV/$c$) & $n$ & $k$ & $\langle p_{Ti} \rangle$ (GeV/$c$) & $N_{p_T}$ & $\chi^2$/dof \\
\hline
Figure 1(c) & hadronic $t$ & $225.00\pm22.50$ & $6.50\pm1.00$ & $0.10\pm0.02$ & $39.70\pm0.50$ & $(1.99\pm0.05)\times10^{5}$ & 11.34 \\
Figure 2(c) & hadronic $t$ & $204.00\pm20.40$ & $6.05\pm1.00$ & $0.13\pm0.02$ & $40.50\pm0.05$ & $1.00$ & 14.33 \\
Figure 3(c) & $t$          & $165.00\pm16.50$ & $5.60\pm1.00$ & $0.14\pm0.02$ & $39.20\pm0.05$ & $1.00$ & 3.76 \\
\hline
\end{tabular}%
\end{center}}
\end{table*}

\begin{figure*}
\hskip-1.0cm \begin{center}
\includegraphics[width=15.0cm]{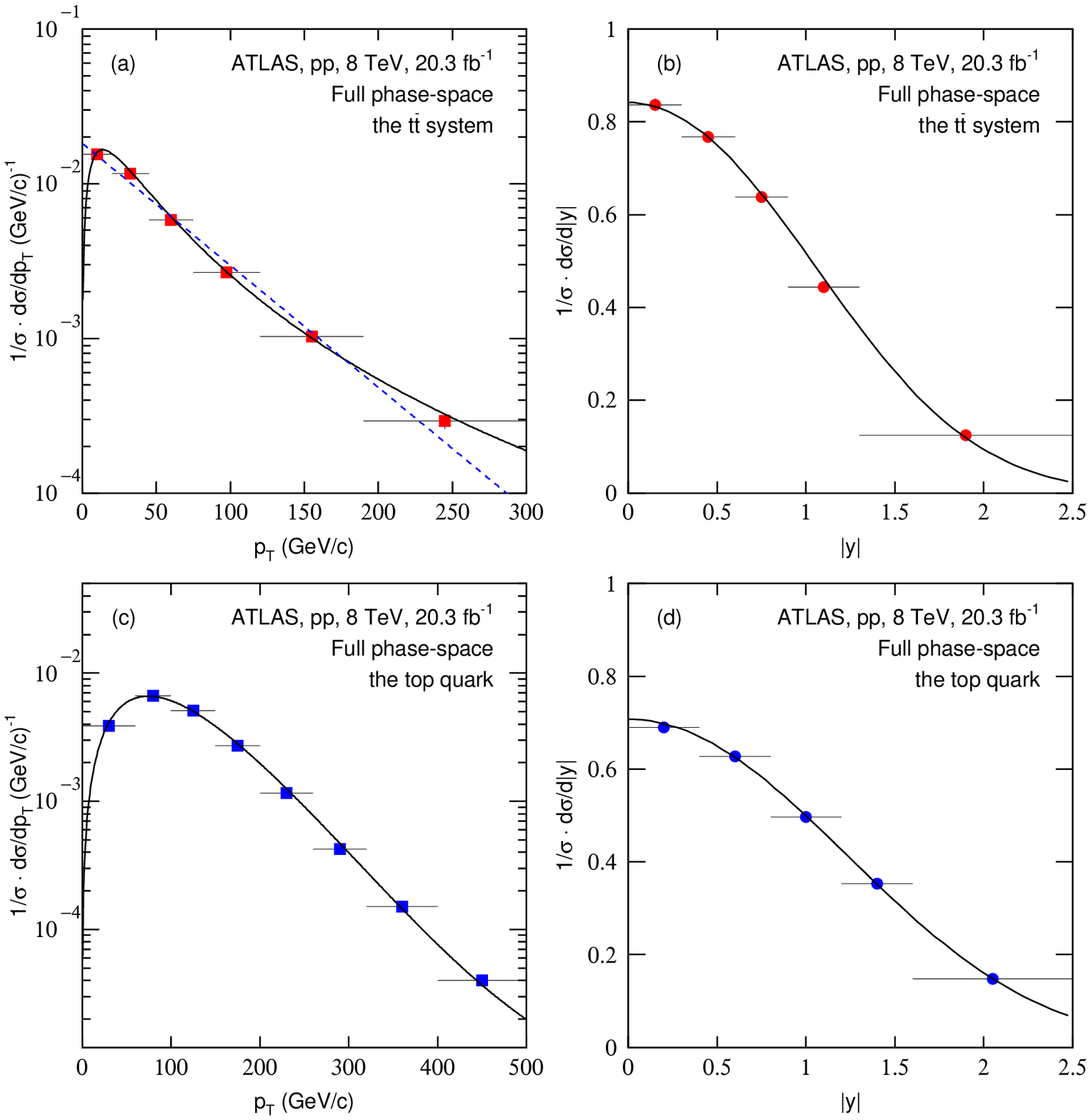}
\end{center}
\vskip0.5cm {\small Fig. 3. (a)(c) Transverse momentum and (b)(d)
rapidity normalized differential cross-sections of (a)(b) the
$t\bar t$ systems and (c)(d) the top quarks produced in $pp$
collisions at $\sqrt{s}=8$ TeV. The symbols represent the
experimental data of the ATLAS Collaboration [30] measured in the
combined electron and muon selections at the full phase-space
level. The solid curves are our results calculated by using the
(a) inverse power-law, (c) superposition of inverse power-law and
Erlang distribution, and (b)(d) two-component Gaussian
distribution, respectively. The dashed curve in (a) is the result
calculated by using the Erlang distribution with $m=1$, which is
in fact the exponential distribution.}
\end{figure*}

\begin{table*}
{\scriptsize Table 3. Values of free parameter [$y_F$ ($=-y_B$)
and $\sigma_{y_F}$ ($=\sigma_{y_B}$)], normalization constant
($N_\eta$), and $\chi^2$/dof corresponding to the curves in
Figures 1(b), 1(d), 2(b), 2(d), 3(b), and 3(d).
\begin{center}
\begin{tabular}{cccccc}
\hline\hline  Figure & Type & $y_F$ ($=-y_B$) & $\sigma_{y_F}$ ($=\sigma_{y_B}$) & $N_{y}$ & $\chi^2$/dof \\
\hline
Figure 1(b) & $t\bar t$    & $0.36\pm0.02$ & $0.53\pm0.02$ & $(196.00\pm2.00)\times10^3$ & 2.27 \\
Figure 1(d) & hadronic $t$ & $0.53\pm0.02$ & $0.68\pm0.02$ & $(199.00\pm2.00)\times10^3$ & 2.21 \\
Figure 2(b) & $t\bar t$    & $0.39\pm0.02$ & $0.55\pm0.02$ & $1.00$                      & 7.88 \\
Figure 2(d) & hadronic $t$ & $0.57\pm0.02$ & $0.69\pm0.02$ & $1.01\pm0.01$               & 13.46\\
Figure 3(b) & $t\bar t$    & $0.52\pm0.02$ & $0.79\pm0.03$ & $1.04\pm0.01$               & 4.64 \\
Figure 3(d) & $t$          & $0.58\pm0.02$ & $0.98\pm0.03$ & $1.04\pm0.01$               & 6.55 \\
\hline
\end{tabular}%
\end{center}}
\end{table*}

\begin{table*}
{\scriptsize Table 4. Values of the root-mean-squares
$\sqrt{\overline{\beta_x^2}}$ for $\beta_x$,
$\sqrt{\overline{\beta_y^2}}$ for $\beta_y$, and
$\sqrt{\overline{\beta_z^2}}$ for $\beta_z$, as well as the
maximum $|\beta_x|$, $|\beta_y|$, and $|\beta_z|$
($|\beta_x|_{\max}$, $|\beta_y|_{\max}$, and $|\beta_z|_{\max}$)
corresponding to the scatter plots for different types of
products, where the corresponding scatter plots are presented in
Fig. 4. Both the root-mean-squares and the maximum velocity
components are in units of $c$.
\begin{center}
\begin{tabular}{cccccccc}
\hline\hline Figure & Type & $\sqrt{\overline{\beta_x^2}}$ & $\sqrt{\overline{\beta_y^2}}$ & $\sqrt{\overline{\beta_z^2}}$ & $|\beta_x|_{\max}$ & $|\beta_y|_{\max}$ & $|\beta_z|_{\max}$ \\
\hline
Figure 4(a) & $t\bar t$    & $0.11\pm0.01$ & $0.11\pm0.01$ & $0.50\pm0.01$ & 0.58 & 0.56 & 0.96 \\
Figure 4(b) & hadronic $t$ & $0.31\pm0.01$ & $0.32\pm0.01$ & $0.59\pm0.01$ & 0.79 & 0.82 & 0.99 \\
Figure 4(c) & $t\bar t$    & $0.13\pm0.01$ & $0.14\pm0.01$ & $0.52\pm0.01$ & 0.76 & 0.67 & 0.97 \\
Figure 4(d) & hadronic $t$ & $0.31\pm0.01$ & $0.32\pm0.01$ & $0.61\pm0.01$ & 0.81 & 0.87 & 0.99 \\
Figure 4(e) & $t\bar t$    & $0.10\pm0.01$ & $0.10\pm0.01$ & $0.62\pm0.01$ & 0.52 & 0.46 & 0.99 \\
Figure 4(f) & $t$          & $0.28\pm0.01$ & $0.29\pm0.01$ & $0.67\pm0.01$ & 0.77 & 0.84 & 1.00 \\
\hline
\end{tabular}%
\end{center}}
\end{table*}

As for the tendencies of the free parameters, one can see from
Tables 1--3 that, for both the $t\bar t$ systems and the
(hadronic) top quarks, $p_0$ and $n$ decrease when the
experimental requirement changes from the detector level to the
fiducial phase-space and then to the full phase-space. Only for
the (hadronic) top quarks, $k$ slightly increases, and there is
almost no change in $\langle p_{Ti} \rangle$ and $m$  when the
requirement changes from the detector level to the full
phase-space. At the same time, for both the $t\bar t$ systems and
the (hadronic) top quarks, both $y_F$ and $\sigma_{y_F}$ increase
when the requirement changes from the detector level to the full
phase-space. These tendencies may have no obvious meaning due to
there being little relation among these requirements. However,
because of these tendencies, we can obtain abundant structures of
event patterns.

Based on the parameter values obtained from Figs. 1--3 and
listed in Tables 1--3, Monte Carlo calculation can be
performed and the values of a series of kinematical quantities can
be obtained. Thus, we can get different kinds of diagrammatic
sketches at the last stage of particle production in the
interacting system formed in $pp$ collisions. Figures 4--6 give
the event patterns which are displayed by the particle scatter
plots in the three-dimensional $\beta_x$-$\beta_y$-$\beta_z$,
$p_x$-$p_y$-$p_z$, and $y_1$-$y_2$-$y$ spaces, respectively. In
these figures, panels (a)--(f) correspond to the results for the
$t\bar t$ systems at the detector level, the hadronic top quarks
at the detector level, the $t\bar t$ systems in the fiducial
phase-space, the hadronic top quarks in the fiducial phase-space,
the $t\bar t$ systems in the full phase-space, and the top quarks
in the full phase-space, respectively. The total number of
particles for each panel is 1000. The blue and red globules
represent the contributions of the inverse power-law and Erlang
distribution respectively. The values of root-mean-squares
$\sqrt{\overline{\beta_x^2}}$ for $\beta_x$,
$\sqrt{\overline{\beta_y^2}}$ for $\beta_y$, and
$\sqrt{\overline{\beta_z^2}}$ for $\beta_z$, as well as the
maximum $|\beta_x|$, $|\beta_y|$, and $|\beta_z|$ (i.e.
$|\beta_x|_{\max}$, $|\beta_y|_{\max}$, and $|\beta_z|_{\max}$)
are listed in Table 4. The values of root-mean-squares
$\sqrt{\overline{p_x^2}}$ for $p_x$, $\sqrt{\overline{p_y^2}}$ for
$p_y$, and $\sqrt{\overline{p_z^2}}$ for $p_z$, as well as the
maximum $|p_x|$, $|p_y|$, and $|p_z|$ (i.e. $|p_x|_{\max}$,
$|p_y|_{\max}$, and $|p_z|_{\max}$) are listed in Table 5. The
values of root-mean-squares $\sqrt{\overline{y_1^2}}$ for $y_1$,
$\sqrt{\overline{y_2^2}}$ for $y_2$, and $\sqrt{\overline{y^2}}$
for $y$, as well as the maximum $|y_1|$, $|y_2|$, and $|y|$ (i.e.
$|y_1|_{\max}$, $|y_2|_{\max}$, and $|y|_{\max}$) are listed in
Table 6.

From Figs. 4--6 and Tables 4--6, one can see that the event
patterns in the three-dimensional $\beta_x$-$\beta_y$-$\beta_z$
space for the $t\bar t$ systems in the three requirements are
rough cylinders with $\sqrt{\overline{\beta_x^2}} \approx
\sqrt{\overline{\beta_y^2}} \ll \sqrt{\overline{\beta_z^2}}$ and
$|\beta_x|_{\max} \approx |\beta_y|_{\max} < |\beta_z|_{\max}$,
though few differences among the three requirements are observed.
The event patterns for the (hadronic) top quarks in the three
requirements are rough ellipsoids, with similar relations among
these quantities and few differences between the three
requirements. An obvious difference between the event patterns for
the $t\bar t$ systems and the (hadronic) top quarks is observed
due to their different production processes. Meanwhile, both the
root-mean-squares and the maxima for the $t\bar t$ systems are
less than those for the (hadronic) top quarks, and the differences
in relative sizes between transverse and longitudinal quantities
for the $t\bar t$ systems are larger than those for the (hadronic)
top quarks.

The event patterns in the three-dimensional $p_x$-$p_y$-$p_z$
space for the $t\bar t$ systems in the three requirements are
relatively thin and very rough ellipsoids with
$\sqrt{\overline{p_x^2}} \approx \sqrt{\overline{p_y^2}} \ll
\sqrt{\overline{p_z^2}}$ and $|p_x|_{\max} \approx |p_y|_{\max}
\ll |p_z|_{\max}$, though few differences among the three
requirements are observed. The event patterns for the (hadronic)
top quarks in the three requirements are relatively fat and very
rough ellipsoids with similar relations among these quantities
and few differences between the three requirements. An obvious
difference between the event patterns for the $t\bar t$ systems
and the (hadronic) top quarks is observed. Meanwhile, the
transverse quantities for the $t\bar t$ systems are less than
those for the (hadronic) top quarks, while the opposite is true for the
longitudinal quantities. The differences in relative
sizes between transverse and longitudinal quantities for the
$t\bar t$ systems are larger than those for the (hadronic) top
quarks. The maximum quantities do not show an obvious tendency for
the $t\bar t$ systems and the (hadronic) top quarks.

The event patterns in the three-dimensional $y_1$-$y_2$-$y$ space
for the $t\bar t$ systems in the three requirements are very rough
ellipsoids with $\sqrt{\overline{y_1^2}} \approx
\sqrt{\overline{y_2^2}} \ll \sqrt{\overline{y^2}}$ and
$|y_1|_{\max} \approx |y_2|_{\max} \ll |y|_{\max}$, though few
differences among the three requirements are observed. The event
patterns for the (hadronic) top quarks in the three requirements
are very rough rhomboids with similar relations among these
quantities and few differences between the three requirements. An
obvious difference between the event patterns for the $t\bar t$
systems and the (hadronic) top quarks is observed. Meanwhile,
both the root-mean-squares and the maxima for
the $t\bar t$ systems are obviously less than those for the
(hadronic) top quarks. The differences in relative sizes between
transverse and longitudinal quantities for the $t\bar t$ systems
are larger than those for the (hadronic) top quarks.

\begin{figure*}
\hskip-1.0cm \begin{center}
\includegraphics[width=15.0cm]{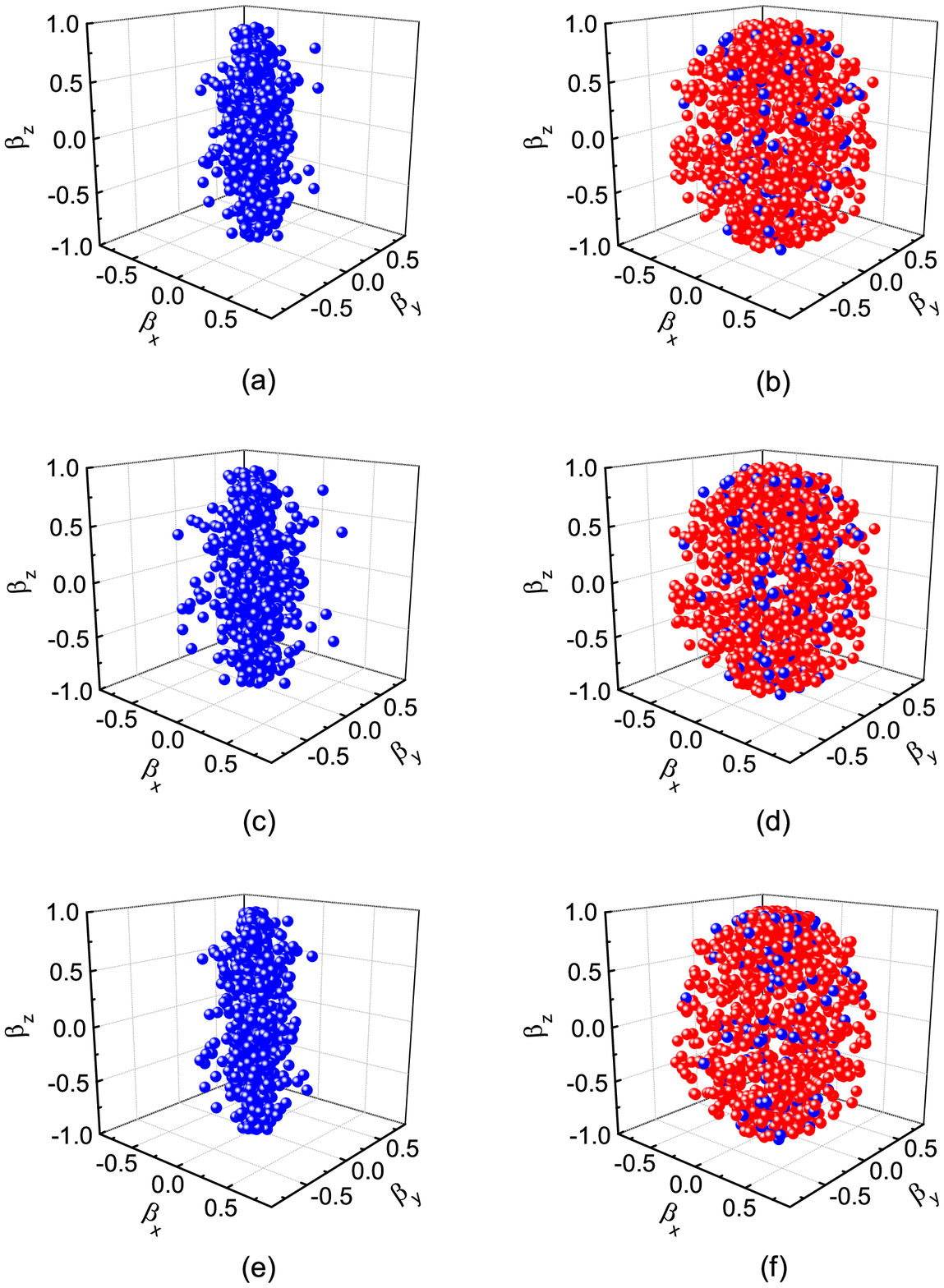}
\end{center}
\vskip0.5cm {\small Fig. 4. Event patterns (particle scatter
plots) in three-dimensional $\beta_x$-$\beta_y$-$\beta_z$ space in
$pp$ collisions at $\sqrt{s}=8$ TeV (a)(b) at the detector level,
(c)(d) in the fiducial phase-space, and (e)(f) in the full
phase-space, for (a)(c)(e) the $t\bar t$ systems, (b)(d) the
hadronic top quarks, and (f) the top quarks. The number of
particles for each panel is 1000. The blue and red globules
represent the results corresponding to the inverse power-law
function and the Erlang distribution for $p_T$, respectively.}
\end{figure*}

\begin{figure*}
\hskip-1.0cm \begin{center}
\includegraphics[width=16.0cm]{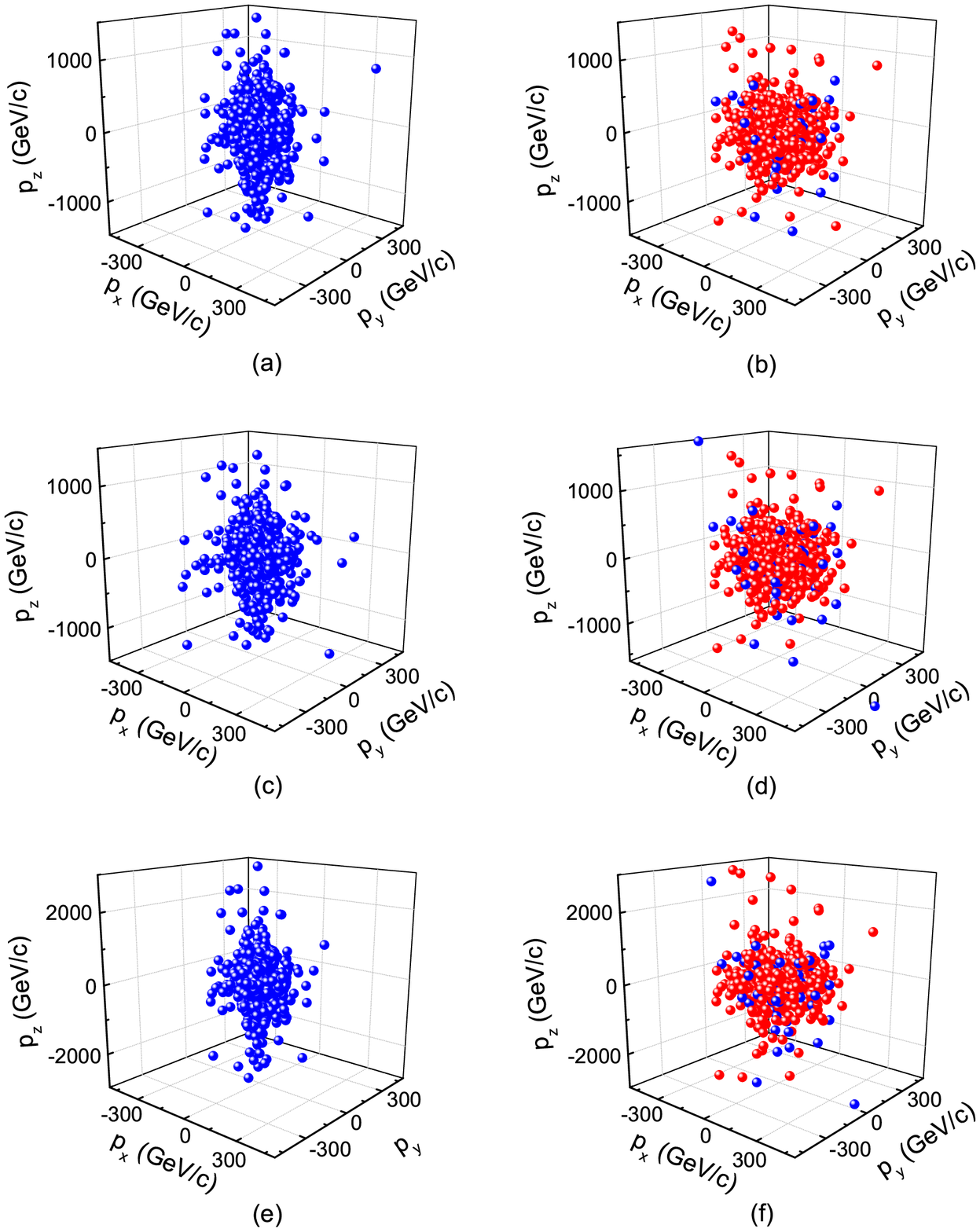}
\end{center}
\vskip0.5cm {\small Fig. 5.
Event patterns (particle scatter
plots) in three-dimensional $p_x$-$p_y$-$p_z$ space in
$pp$ collisions at $\sqrt{s}=8$ TeV (a)(b) at the detector level,
(c)(d) in the fiducial phase-space, and (e)(f) in the full
phase-space, for (a)(c)(e) the $t\bar t$ systems, (b)(d) the
hadronic top quarks, and (f) the top quarks. The number of
particles for each panel is 1000. The blue and red globules
represent the results corresponding to the inverse power-law
function and the Erlang distribution for $p_T$, respectively.
}
\end{figure*}

\begin{figure*}
\hskip-1.0cm \begin{center}
\includegraphics[width=15.0cm]{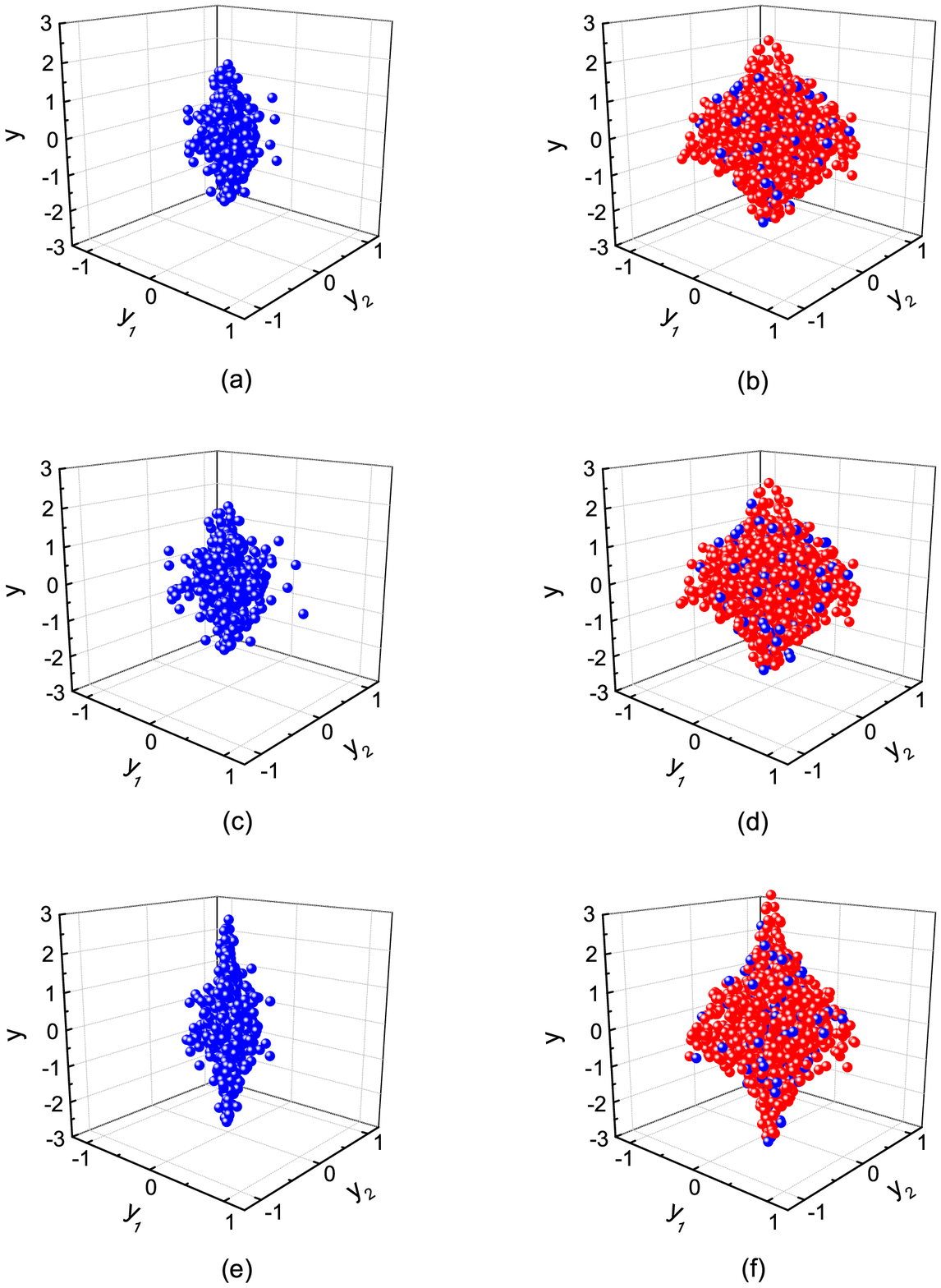}
\end{center}
\vskip0.5cm {\small Fig. 6.
Event patterns (particle scatter
plots) in three-dimensional $y_1$-$y_2$-$y$ space in
$pp$ collisions at $\sqrt{s}=8$ TeV (a)(b) at the detector level,
(c)(d) in the fiducial phase-space, and (e)(f) in the full
phase-space, for (a)(c)(e) the $t\bar t$ systems, (b)(d) the
hadronic top quarks, and (f) the top quarks. The number of
particles for each panel is 1000. The blue and red globules
represent the results corresponding to the inverse power-law
function and the Erlang distribution for $p_T$, respectively.
}
\end{figure*}

\begin{figure*}
\hskip-1.0cm \begin{center}
\includegraphics[width=15.0cm]{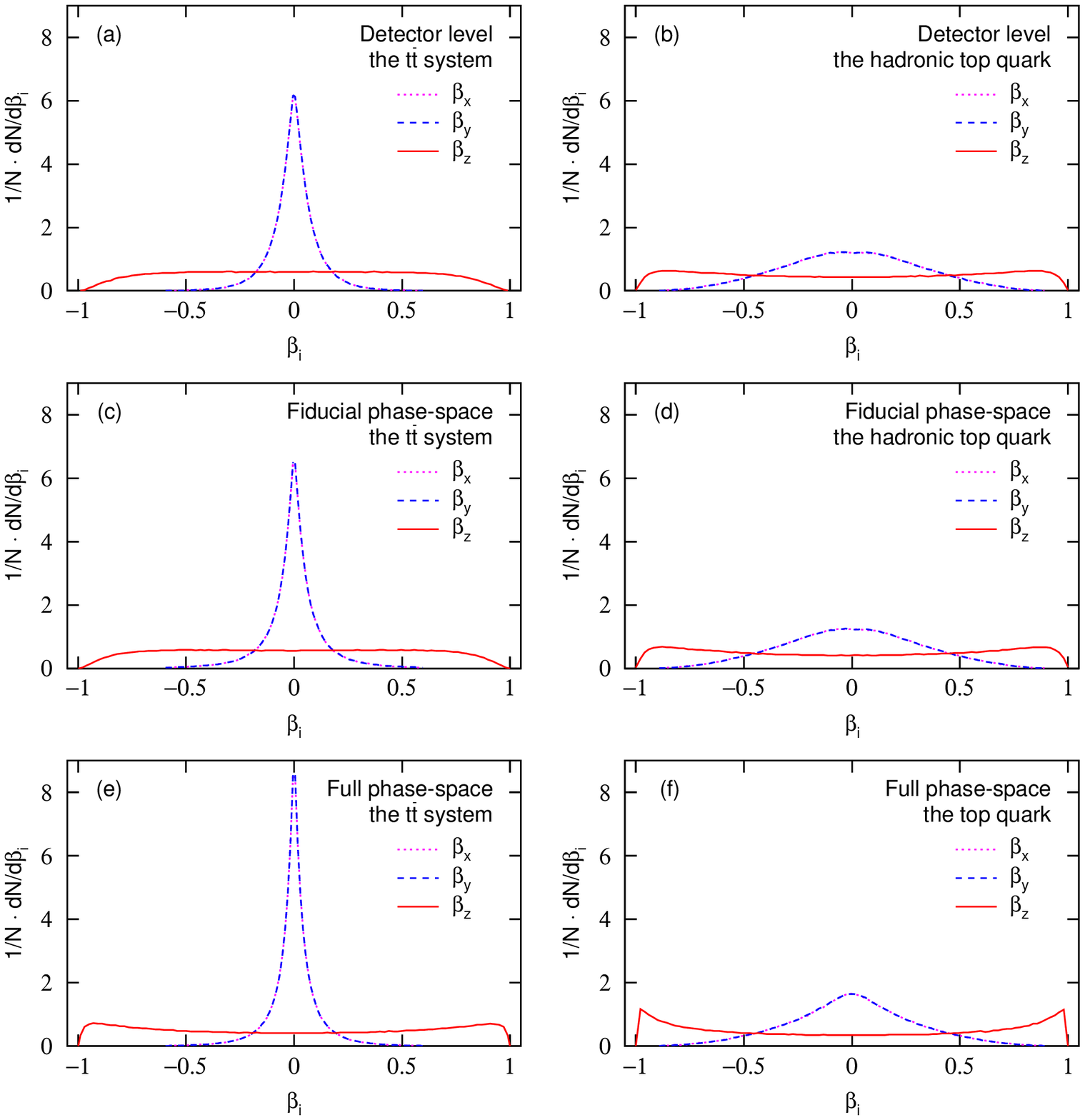}
\end{center}
\vskip0.5cm {\small Fig. 7. Distributions of velocity components
$\beta_x$, $\beta_y$, and $\beta_z$ in $pp$ collisions at
$\sqrt{s}=8$ TeV (a)(b) at the detector level, (c)(d) in the
fiducial phase-space, and (e)(f) in the full phase-space, for
(a)(c)(e) the $t\bar t$ systems, (b)(d) the hadronic top quarks,
and (f) the top quarks. The dotted, dashed, and solid curves
correspond to the distributions of $\beta_x$, $\beta_y$, and
$\beta_z$, respectively, where the distributions of $\beta_x$ and
$\beta_y$ are nearly the same.}
\end{figure*}

\begin{figure*}
\hskip-1.0cm \begin{center}
\includegraphics[width=15.0cm]{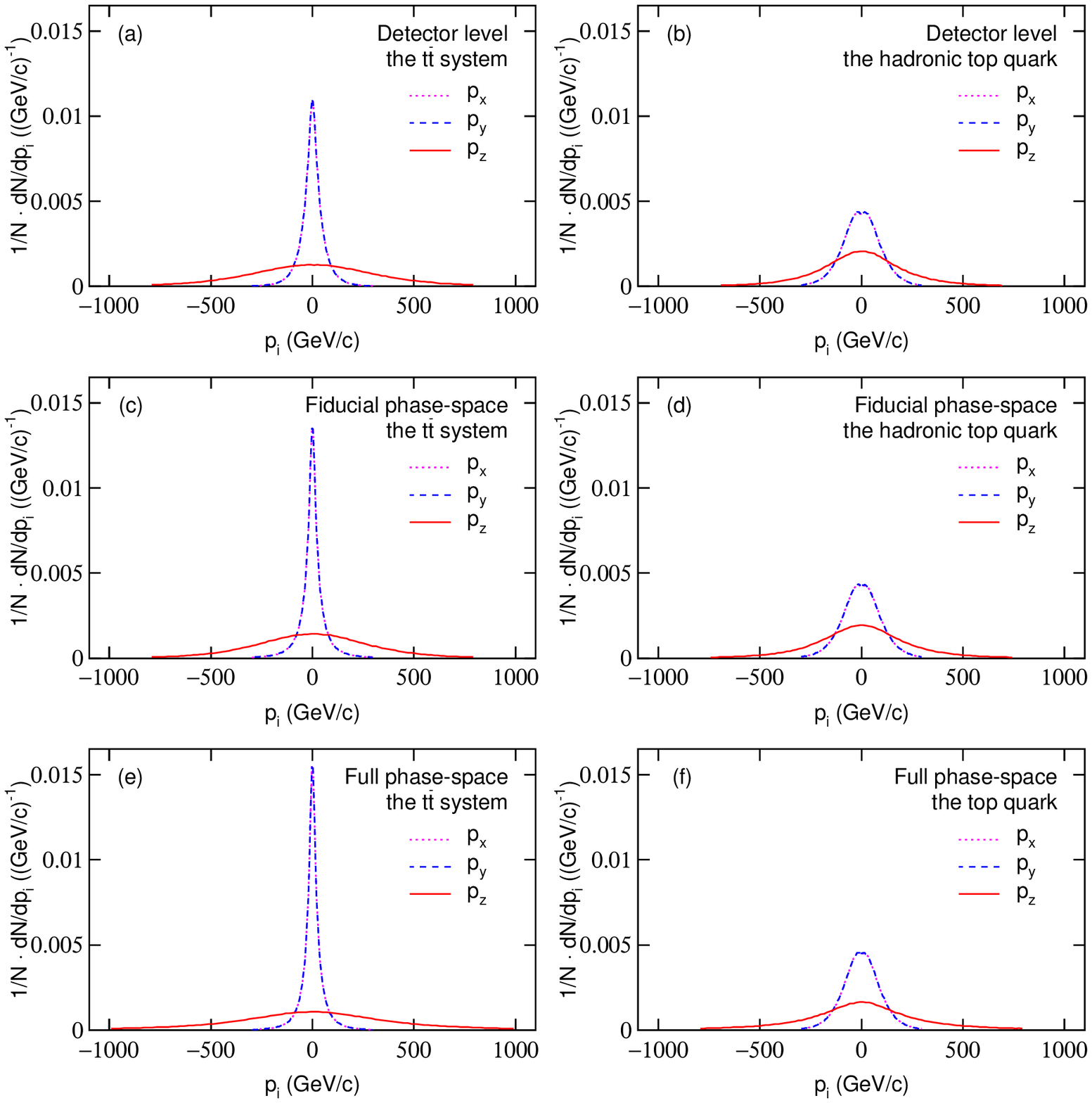}
\end{center}
\vskip0.5cm {\small Fig. 8. Distributions of momentum components
$p_x$, $p_y$, and $p_z$ in $pp$ collisions at $\sqrt{s}=8$ TeV
(a)(b) at the detector level, (c)(d) in the fiducial phase-space,
and (e)(f) in the full phase-space, for (a)(c)(e) the $t\bar t$
systems, (b)(d) the hadronic top quarks, and (f) the top quarks.
The dotted, dashed, and solid curves correspond to the
distributions of $p_x$, $p_y$, and $p_z$, respectively, where the
distributions of $p_x$ and $p_y$ are nearly the same.}
\end{figure*}

\begin{figure*}
\hskip-1.0cm \begin{center}
\includegraphics[width=15.0cm]{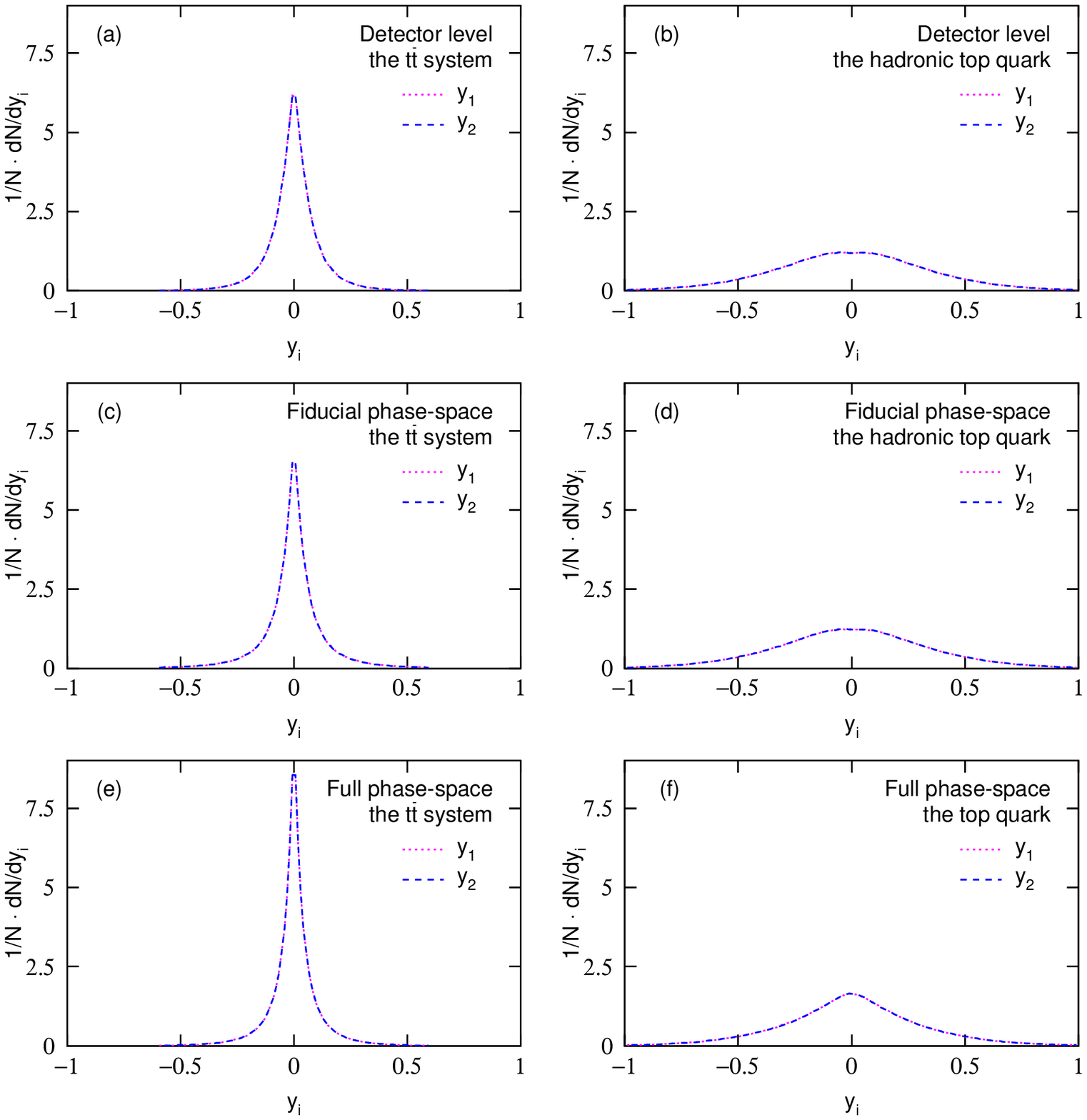}
\end{center}
\vskip0.5cm {\small Fig. 9. Distributions of $y_1$ and $y_2$ in
$pp$ collisions at $\sqrt{s}=8$ TeV (a)(b) at the detector level,
(c)(d) in the fiducial phase-space, and (e)(f) in the full
phase-space, for (a)(c)(e) the $t\bar t$ systems, (b)(d) the
hadronic top quarks, and (f) the top quarks. The dotted and dashed
curves correspond to the distributions of $y_1$ and $y_2$
respectively, where the two distributions are nearly the same.}
\end{figure*}

\begin{table*}
{\scriptsize Table 5. Values of the root-mean-squares
$\sqrt{\overline{p_x^2}}$ for $p_x$, $\sqrt{\overline{p_y^2}}$ for
$p_y$, and $\sqrt{\overline{p_z^2}}$ for $p_z$, as well as the
maximum $|p_x|$, $|p_y|$, and $|p_z|$ ($|p_x|_{\max}$,
$|p_y|_{\max}$, and $|p_z|_{\max}$) corresponding to the scatter
plots for different types of products, where the corresponding
scatter plots are presented in Fig. 5. Both the
root-mean-squares and maximum momentum components are in units
of GeV/$c$.
\begin{center}
\begin{tabular}{cccccccc}
\hline\hline Figure & Type & $\sqrt{\overline{p_x^2}}$  & $\sqrt{\overline{p_y^2}}$  & $\sqrt{\overline{p_z^2}}$  & $|p_x|_{\max}$  & $|p_y|_{\max}$  & $|p_z|_{\max}$  \\
\hline
Figure 5(a) & $t\bar t$    & $63.2\pm2.8$ & $63.9\pm2.8$ & $373.4\pm11.0$  & 427.4 & 469.0 & 1545.0 \\
Figure 5(b) & hadronic $t$ & $97.0\pm2.7$ & $98.8\pm2.5$ & $279.1\pm10.8$  & 433.6 & 384.8 & 1347.2 \\
Figure 5(c) & $t\bar t$    & $75.1\pm4.7$ & $74.9\pm4.4$ & $338.2\pm10.3$  & 589.3 & 632.2 & 1399.1 \\
Figure 5(d) & hadronic $t$ & $100.9\pm3.1$& $102.8\pm3.0$& $304.9\pm13.4$  & 498.6 & 560.5 & 1999.9 \\
Figure 5(e) & $t\bar t$    & $52.4\pm2.1$ & $53.2\pm2.0$ & $574.6\pm23.7$  & 273.5 & 255.6 & 3189.3 \\
Figure 5(f) & $t$          & $95.0\pm2.6$ & $96.8\pm2.5$ & $513.9\pm28.8$  & 393.8 & 423.9 & 3079.7 \\
\hline
\end{tabular}%
\end{center}}
\end{table*}

\begin{table*}
{\scriptsize Table 6. Values of the root-mean-squares
$\sqrt{\overline{y_1^2}}$ for $y_1$, $\sqrt{\overline{y_2^2}}$ for
$y_2$, and $\sqrt{\overline{y^2}}$ for $y$, as well as the maximum
$|y_1|$, $|y_2|$, and $|y|$ ($|y_1|_{\max}$, $|y_2|_{\max}$, and
$|y|_{\max}$) corresponding to the scatter plots for different
types of products, where the corresponding scatter plots are
presented in Fig. 6.
\begin{center}
\begin{tabular}{cccccccc}
\hline\hline Figure & Type & $\sqrt{\overline{y_1^2}}$  & $\sqrt{\overline{y_2^2}}$  & $\sqrt{\overline{y^2}}$  & $|y_1|_{\max}$  & $|y_2|_{\max}$ & $|y|_{\max}$ \\
\hline
Figure 6(a) & $t\bar t$    & $0.11\pm0.01$ & $0.11\pm0.01$ & $0.65\pm0.01$ & 0.67 & 0.63 & 1.92 \\
Figure 6(b) & hadronic $t$ & $0.34\pm0.01$ & $0.36\pm0.01$ & $0.87\pm0.02$ & 1.08 & 1.16 & 2.53 \\
Figure 6(c) & $t\bar t$    & $0.14\pm0.01$ & $0.15\pm0.01$ & $0.68\pm0.01$ & 1.01 & 0.81 & 2.01 \\
Figure 6(d) & hadronic $t$ & $0.34\pm0.01$ & $0.36\pm0.01$ & $0.90\pm0.02$ & 1.12 & 1.34 & 2.60 \\
Figure 6(e) & $t\bar t$    & $0.10\pm0.01$ & $0.11\pm0.01$ & $0.95\pm0.02$ & 0.58 & 0.49 & 2.84 \\
Figure 6(f) & $t$          & $0.30\pm0.01$ & $0.33\pm0.01$ & $1.15\pm0.03$ & 1.02 & 1.22 & 3.46 \\
\hline
\end{tabular}%
\end{center}}
\end{table*}

According to these scatter plots (Figs. 4--6), we can obtain the
probability distributions of the considered quantities. Using
higher statistics, Figs. 7--9 present the probability
distributions of $\beta_i$ ($i=x$, $y$, and $z$), $p_i$ ($i=x$,
$y$, and $z$), and $y_i$ ($i=1$ and 2), respectively, where $N$
denotes the number of particles. Different panels correspond to
different requirements and different curves correspond to
different quantities shown in the panels. One can see that the
distributions of $x$ and $y$ components are almost the same, if
not equal to each other at the pixel level, due to the assumption
of isotropic emission in the transverse plane. All distributions
of $x$, $y$ and $z$ components are symmetric at zero.

On the velocity components, $\beta_x$ and $\beta_y $ are distributed
only in a small region near zero. The $t\bar t$ systems have a
narrower region and a higher peak than the (hadronic) top
quarks. For the requirements from the detector level to the
fiducial phase-space then to the full phase-space, the peak value
increases obviously. $\beta_z$ is distributed almost uniformly in a
wide region for different particles and requirements. In
particular, the distributions of $|\beta_z|$ of the (hadronic) top
quarks increase slightly with the increase of $|\beta_z|$, while
the $t\bar t$ systems show an opposite or static tendency.
Moreover, the tendency in the full phase-space is more obvious
than in the fiducial phase-space and at the detector level.
The main differences appear near $|\beta_z|_{\max}$ and are caused
by different $m_0$.

On the momentum components, all the distributions of $p_x$, $p_y$,
and $p_z$ for different particles and requirements have a peak at
zero, though the distribution of $p_z$ has a wider range and a
lower peak than those of $p_x$ and $p_y$. The $t\bar t$ system
has an increasingly high peak in the $p_x$ and $p_y$ distributions  at the detector level, in the fiducial phase-space,
and in the full phase-space, while the other three types of
distributions have similar shapes and do not show an
obvious tendency to increasing peak height.

For the distributions of $y_1$ and $y_2$, one can see an increasingly high peak for the $t\bar t$ system going
from the detector level to the fiducial phase-space then to the
full phase-space. For the hadronic top quarks, the distributions
of $y_1$ and $y_2$ at the detector level and in the fiducial
phase-space are almost the same, and they are different in shape
and slope around the peak region from the distributions for the
top quarks.

It should be noted that, in the above discussions, we have
produced fits to the ATLAS 8 TeV data on $p_T$ and $y$ of the top
quarks and the $t\bar t$ pairs. The fits for these one-dimensional
kinematic distributions are then used in such a way that fully
differential distributions for the same final states are
``predicted". It seems that, in general, it is impossible to
achieve this because one cannot reconstruct a generic
multidimensional distribution from its one-dimensional (marginal)
projections due to the correlations between variables not being
known. In fact, our procedure of reconstruction works well due to
the correlations between variables being considered through Eqs.
(11) and (14).

For the correlations between $x$- and $y$-components, we have
used an isotropic assumption in the transverse plane. This results
in Eqs. (10)--(12) for the azimuthal angle, $p_x$, and $p_y$,
respectively. We would like to point out that the effect of
elliptic flow for the top quarks and the $t\bar t$ pairs are
neglected due to this effect appearing mainly in the soft process.
In the case of considering the elliptic flow for the soft process,
we are expected to study the fine-structure of event patterns
[31], which is beyond the focus of the present work, though this
effect is small and can be neglected. In any case,
conservation of energy and momentum is satisfied in the
calculation.

For comparisons with our recent works [23--25], as an example, we
can see the similarities and differences in the three-dimensional
$\beta_x$-$\beta_y$-$\beta_z$ space. The scatter plots of $t\bar
t$ systems and (hadronic) top quarks are similar to those of $Z$
bosons and quarkonium states [25] due to them being heavy
particles. In fact, the scatter plots of heavy particles show that
the root-mean square velocities form a rough cylinder or ellipsoid
surface and the maximum velocities form a fat cylinder or
ellipsoid surface, due to their production being at the initial
stage of collisions. The scatter plots of charged particles show
that the root-mean-square velocities form an ellipsoid surface and
the maximum velocities form a spherical surface [23, 24], due to
their production being mostly at the intermediate stage of
collisions and suffering particularly the processes of
thermalization and expansion of the interacting system.

As for comparisons with other modelling or theoretical works,
although thousands of papers on top quark-related subjects
have been published since (at least) the 1980s, and the number of
top quark-related publications from the LHC is also in the
hundreds (for example, see Refs.~[32--36]), few of them are directly
related to the event patterns or particle scatter plots. In fact,
we cannot give a direct comparison with other works due to the
available results not being obtained. In addition, although one
might just use some event generators such as Pythia, JETSET, and
HERWIG instead [37--41], they are not just fitted to transverse
momentum and rapidity spectra and require information about
the underlying event, pileup, and so on. The present work provides
a simple and alternative method to structure event patterns
displayed by the scatter plots of different particles. Using this
alternative method, one can obtain some direct and ikonic pictures
for production of different particles.

Although we have used the hybrid model to fit the experimental
$p_T$ and $y$ spectra to extract the parameter values and to
restructure the event patterns, the event patterns we have obtained
are model-independent. In particular, similar or related
experimental spectra are also described or predicted by other
perturbative QCD calculations such as the Next-to-Leading Order
(NLO) in QCD [42--44], Next-to-Next-to-Leading Order (NNLO) in QCD
[45--50], Next-to-Next-to-Leading Logarithms (NNLL) [51--53], etc.
The event patterns are independent of these theories. In fact, the
event patterns are only dependent on the discrete values of
experimental probability density distributions of $p_T$ and $y$.
What we fitted in the above by using the hybrid model is only
parameterizations for the $p_T$ and $y$ spectra. These
parameterizations smooth only the experimental probability density
distributions and help us to restructure the event patterns.

Before giving conclusions, we would like to emphasize briefly the
significance of the present work. In our opinion, the present work
supports the methodology which restructures the event patterns or
particle scatter plots from both $p_T$ and $y$ spectra. This
method can be used in the studies of other particles [23--25, 31]
as discussed above, which allows us to give comparisons of the
production of different types of particles. Indeed, from the
three-dimensional distribution, we have obviously observed some
differences for different type of particles. These differences are
useful in the understanding of particle production and event
reconstruction.
\\

{\section{Conclusions}}

We summarize here our main observations and conclusions.

(a) We have used the hybrid model to fit the top quark-related
spectra of $p_T$ and $y$, which include the spectra of $t\bar t$
systems, hadronic top quarks, and top quarks produced in $pp$
collisions at $\sqrt{s}=8$ TeV measured by the ATLAS Collaboration
at the LHC. The hybrid model uses the superposition of the inverse
power-law and the Erlang distribution for the description of $p_T$
spectra and the two-component Gaussian function for the
description of $y$ spectra. The inverse power-law, the Erlang
distribution, and the two-component Gaussian function are derived
from the QCD calculus, the multisource thermal model, and the
Landau hydrodynamic model, respectively. We have used the inverse
power-law and the Erlang distribution to fit the harder and hard
scattering processes respectively.

(b) The modelling results are in agreement with the experimental
data of the $t\bar t$ systems and the hadronic top quarks measured
at the detector level, the fiducial phase-space normalized
differential cross-sections for the $t\bar t$ systems and the
hadronic top quarks, and the full phase-space normalized
differential cross-sections for the $t\bar t$ systems and the top
quarks. The $t\bar t$ systems show only the contribution of the harder
scattering process (with $k=1$). The (hadronic) top quarks show
mainly the contribution of the hard scattering process (with a small
$k$). This implies that more collision energy is needed to
create the $t\bar t$ system. As a part of the $t\bar t$ system,
the (hadronic) top quark takes up part of the energy of the $t\bar
t$ system, which results in a not too hard scattering process.

(c) When the experimental requirement changes from the detector
level to the fiducial phase-space and then to the full
phase-space, for both the $t\bar t$ systems and the (hadronic) top
quarks, $p_0$ and $n$ decrease, and $y_F$ and $\sigma_{y_F}$
increase. Only for the (hadronic) top quarks, $k$ slightly
increases, while there is almost no change in $\langle p_{Ti} \rangle$ and $m$. Although these tendencies of the parameters may have no
obvious meaning, due to there being little relation among these experimental
requirements, these parameters can be used in the extraction of
discrete values of some kinematic quantities. In fact, based on
these parameters, we have obtained some discrete values of the
velocity, momentum, and rapidity components. Based on these
discrete values, the event patterns in some three-dimensional
spaces are obtained.

(d) The event patterns in the three-dimensional
$\beta_x$-$\beta_y$-$\beta_z$ space for the $t\bar t$ systems in
the three requirements are rough cylinders with
$\sqrt{\overline{\beta_x^2}} \approx \sqrt{\overline{\beta_y^2}}
\ll \sqrt{\overline{\beta_z^2}}$ and $|\beta_x|_{\max} \approx
|\beta_y|_{\max} < |\beta_z|_{\max}$. The event patterns for the
(hadronic) top quarks in the three requirements are rough
ellipsoids with similar relations among these quantities. Both
the root-mean-squares and the maxima for the
$t\bar t$ systems are less than those for the (hadronic) top
quarks, and the differences in relative sizes between transverse
and longitudinal quantities for the $t\bar t$ systems are larger
than those for the (hadronic) top quarks.

(e) The event patterns in the three-dimensional $p_x$-$p_y$-$p_z$
space for the $t\bar t$ systems in the three requirements are
relatively thin and very rough ellipsoids with
$\sqrt{\overline{p_x^2}} \approx \sqrt{\overline{p_y^2}} \ll
\sqrt{\overline{p_z^2}}$ and $|p_x|_{\max} \approx |p_y|_{\max}
\ll |p_z|_{\max}$. The event patterns for the (hadronic) top
quarks in the three requirements are relatively fat and very rough
ellipsoids with the similar relations among these quantities. The
transverse quantities for the $t\bar t$ systems are less than
those for the (hadronic) top quarks, and the situations of
longitudinal quantities are opposite. The differences in relative
sizes between transverse and longitudinal quantities for the
$t\bar t$ systems are larger than those for the (hadronic) top
quarks. The maximum quantities do not show an obvious tendency for
the $t\bar t$ systems and the (hadronic) top quarks.

(f) The event patterns in the three-dimensional $y_1$-$y_2$-$y$
space for the $t\bar t$ systems in the three requirements are very
rough ellipsoids with $\sqrt{\overline{y_1^2}} \approx
\sqrt{\overline{y_2^2}} \ll \sqrt{\overline{y^2}}$ and
$|y_1|_{\max} \approx |y_2|_{\max} \ll |y|_{\max}$. The event
patterns for the (hadronic) top quarks in the three requirements
are very rough rhomboids with similar relations among these
quantities. Both the root-mean-squares and the
maxima for the $t\bar t$ systems are obviously less than those
for the (hadronic) top quarks. The differences in relative sizes
between transverse and longitudinal quantities for the $t\bar t$
systems are larger than those for the (hadronic) top quarks.

(g) According to these scatter plots, we have obtained the
probability distributions of the considered quantities such as
$\beta_i$, $p_i$, and $y_i$. The distributions of $x$ and $y$
components are almost the same, if not equal to each other at the
pixel level, due to the assumption of isotropic emission in the
transverse plane. $\beta_x$ and $\beta_y $ are distributed only in a
small region near zero. The $t\bar t$ systems have a  narrower
region and a higher peak than the (hadronic) top quarks. $\beta_z$ is
distributed almost uniformly in a wide region for different
particles and requirements. In particular, the distributions of
$|\beta_z|$ of the (hadronic) top quarks increase slightly with
the increase of $|\beta_z|$, while the $t\bar t$ systems show an
opposite or static tendency.

(h) All the distributions of $p_x$, $p_y$, and $p_z$ for different
particles and requirements have a peak at zero, though the
distribution of $p_z$ has a wider range and a lower peak than
those of $p_x$ and $p_y$. The $t\bar t$ systems have a higher peak
in the $p_x$ and $p_y$ distributions, while the other three types of
distributions have similar shapes and do not obviously show
 such a high peak. In the distributions of $y_1$ and
$y_2$, a higher peak for the $t\bar t$ systems than for the
(hadronic) top quarks is observed. For the hadronic top quarks,
the distributions of $y_1$ and $y_2$ at the detector level and in
the fiducial phase-space are almost the same, and they are
different in shape and slope around the peak region from the
distributions for the top quarks.
\\

{\bf Conflict of Interests}

The authors declare that there is no conflict of interests
regarding the publication of this paper.
\\

{\bf Acknowledgments}

This work was supported by the National Natural Science Foundation
of China under Grant Nos. 11575103 and 11747319, the Shanxi
Provincial Natural Science Foundation under Grant No.
201701D121005, the Fund for Shanxi ``1331 Project" Key Subjects
Construction, and the US DOE under contract
DE-FG02-87ER40331.A008.
\\

\end{multicols}
\end{document}